\documentclass{IOS-Book-Article}
\usepackage{times}
\normalfont
\usepackage[T1]{fontenc}
\usepackage{graphicx}

\newcommand{\cat}[1]{\left|#1\right\rangle}
\newcommand{\bra}[1]{\left\langle#1\right|}
\newcommand{\brcat}[2]{\left\langle#1\left|#2\right|#1\right\rangle}
\newcommand{\roa}[1]{\left|#1\right\rangle\odot\left\langle#1\right|}
\newcommand{\ecat}[1]{\left|#1\right)}
\newcommand{\ebra}[1]{\left(#1\right|}

\newcommand{\bracat}[3]{\left\langle#1\left|#2\right|#3\right\rangle}
\newcommand{\pcat}[1]{\left|\left.#1\right\rangle\right\rangle}
\newcommand{\pbra}[1]{\left\langle\left\langle#1\right.\right|}

\newcommand{\aver}[1]{\left\langle #1\right\rangle}
\newcommand{\tr}[1]{Tr{\left( #1\right)}}
\newcommand{\eroa}[1]{\left|#1\right)\odot\left(#1\right|}
\newcommand{\proa}[1]{\left|\left.#1\right\rangle\right\rangle \odot\left\langle\left\langle#1\right.\right|}
\newcommand{\sth}[1]
{-#1\log_2{\left( #1\right)}-\left(1- #1\right)\log_2{\left(1- #1\right)}
}
\newcommand{\sthh}[1]{-#1\log_2{\left( #1\right)}
}
\newcommand{\eref}[1]{Eq. (\ref{#1})}
\begin{document}
\begin{frontmatter}                           

\title{Production of Information and Entropy in measurement of entangled states }
\runningtitle{Production of Information }

\author{Constantin V. Usenko
\thanks{prosp. akad. Glushkova, 2, b.1, Kyiv, Ukraine, e-mail: usenko@univ.kiev.ua}}

\runningauthor{Constantin V. Usenko}
\address{National Taras Shevchenko University of Kyiv, Physics Faculty, Department of Theoretical Physics}

\begin{abstract}
Problem of classification of all the set of entangled states is considered.
Invariance of entangled states relative to transformations from a group of symmetry 
of qubit space leads to 
classification of all states of the system through irreducible representations from that group.

Excess of entropy of a subsystem over entropy of the whole system indicates the presence of entaglement in the system.

\end{abstract}

\end{frontmatter}

\thispagestyle{empty}
\pagestyle{empty}

\section*{Introduction}
One of the most interesting phenomena in quantum physics is the ability of quantum system to create information, for instance, in measurement \cite{VP,wootters89,MUBDWF} of electron spins for an EPR-pair.
This property is actively used in different areas of quantum physics and its applications, like Quantum Key Distribution. States of quantum system with such peculiarity are known as entangled states. Recently a lot of entangled states have been studied and there exists an urgent problem of classification of all the set of entangled states.
  Subject of the talk deals with the idea of the invariance of entangled states relative to transformations from a group of symmetry \cite{symbqs,symiso,symws,Rotation,schliemann}. 
Each state of quantum system is invariant relative to phase coefficient thus composite system is to be invariant relative to transformations from the group of symmetry of each subsystem. These groups form the group of symmetry of the whole system so the set of all states of the system can be clasified through irreducible representations of that group.

In this work it is shown that 
almost each space of irreducible representation consists of the entangled states only. Entropy of substates from each space with nontrivial representation exceeds entropy of whole state.
Excess of entropy of a subsystem over entropy of the whole system indicates the system is entagled.
\section{Measurability of Entanglement}

\subsection{General entangled system}

Let the system $S$ has two parts $\mathcal{A}$ and $\mathcal{B}$ and is prepared in common state $\rho_{sys}$. Of course each part is in its own state: part  $\mathcal{A}$ is in the state $\hat{\rho}_{A}$ and part  $\mathcal{B}$ is in the state $\hat{\rho}_{B}$. They can be combined to $\hat{\rho}_{sys}$ in different ways:
\begin{enumerate}
	\item If $\hat{\rho}_{sys}=\hat{\rho}_{A}\otimes\hat{\rho}_{B}$, parts of the system are indepent.
	\item If $\hat{\rho}_{sys}=\sum{p_k\left(\hat{\rho}_{A,k}\otimes\hat{\rho}_{B,k}\right)}$, the system is a mixture of its parts.
	\item All the other states of the system are entangled.
\end{enumerate}

Common example of entangled state is the EPR (Einstein-Podolsky-Rosen) one which is a singlet state of electron pair. In accordance with the principle of identity this state is a linear superposition of states "`spin-up -- spin-down"' and "`spin-down -- spin-up"'
\[\pcat{EPR}=\frac{1}{\sqrt{2}}
\left(\cat{\uparrow}\otimes\cat{\downarrow}-\cat{\downarrow}\otimes\cat{\uparrow}\right).
\]
Another example is Schr\"odinger Cat state being a linear superposition of a photon pair with same polarisations
\[\pcat{Cat}=\frac{1}{\sqrt{2}}
\left(\cat{\updownarrow}\otimes\cat{\updownarrow}
+\cat{\leftrightarrow}\otimes\cat{\leftrightarrow}\right).
\]

\subsubsection{Unitary symmetry}

 Schr\"odinger Cat state shows special type of unitary symmetry.
 
First we denote as $\ecat{0}_A$ and $\ecat{1}_A$ transformed basis of subsystem $\mathcal{A}$: 
\[\cat{0}_A=\cos\theta\ecat{0}_A+e^{i\phi}\sin\theta\ecat{1}_A;\ 
\cat{1}_A=-e^{-i\phi}\sin\theta\ecat{0}_A+\cos\theta\ecat{1}_A.
\]
If basis of subsystem $\mathcal{B}$ is transformed to $\ecat{0}_B$ and $\ecat{1}_B$ by
\[\cat{0}_B=\cos\theta\ecat{0}_B+e^{-i\phi}\sin\theta\ecat{1}_B;\ 
\cat{1}_B=-e^{i\phi}\sin\theta\ecat{0}_A+\cos\theta\ecat{1}_B,
\]
Schr\"odinger Cat remains non-transformed
\[
\frac{1}{\sqrt{2}}\left(\cat{0}_A\otimes\cat{0}_B+\cat{1}_A\otimes\cat{1}_B\right)
=\frac{1}{\sqrt{2}}\left(\ecat{0}_A\otimes\ecat{0}_B+\ecat{1}_A\otimes\ecat{1}_B\right).
\]
Thus, the Schr\"odinger Cat state has group of symmetry $U(2)$ -- group of unitary transformations of two-dimensional space of states.

Similarly, EPR-state has the same group of symmetry.

Systems having larger subsystems can have entangled states with larger group of symmetry $U(N>2)$ but each such group includes $U(2)$ as subgroup, so 

\emph{invariance to group $U(2)$ is essential property of entangled state.}

\subsubsection{States of subsystem}

Now we describe the state of a part of the system (or system as a whole). Under supposition that space of states has finite dimension we describe a state by density matrix
\begin{equation}\label{rho.def}
	\hat{\rho}=\sum_{n,m=1}^{N}{\rho_{n,m}\cat{n}\odot\bra{m}};\ 1\ge\rho_{1,1}\ge\ldots\ge\rho_{N,N}\ge 0.
\end{equation}
As a result of finite dimension we have the solution of eigenvalue problem for the density matrix.
In the case if basis is composed of eigenvectors of density matrix, $\rho_{n\neq m}=0$, we can describe the density matrix as polynomial function $\hat{\rho}=\rho\left(\hat{S}_z\right)$ of momentum operator 
\begin{equation}\label{def_z}
\begin{array}{l}
	\hat{S}_z=\sum_{n=1}^{N}{\left(n-\frac{N+1}{2}\right)\cat{n}\odot\bra{n}};\\
		\hat{S}_+=\sum_{n=1}^{N}{\sqrt{n\left(N-n\right)}\cat{n+1}\odot\bra{n}}
		; \\ \hat{S}_-=\sum_{n=1}^{N}{\sqrt{\left(n-1\right)\left(N-n+1\right)}\cat{n-1}\odot\bra{n}};\\
	\left[\hat{S}_+\hat{S}_-\right]=2\hat{S}_z.
\end{array}
\end{equation}
Now we involve into consideration associate ladder operators $\hat{S}_\pm$ because arbitrary operator on space of states has representation as polynom over ladder operators
\begin{equation}
	\hat{O}=\sum_{m,n=0}^{N-1}{O_{m,n}\hat{S}_+^m\hat{S}_-^n}.
\end{equation}
More exactly each space of subsystem states, as well as the space of states of whole system, is unitarily equivalent to space of irreducible representation of angular momentum $j=\frac{N-1}{2}$.

\subsubsection{States of the composite system}

Space of states of the system is direct multiplication 
$H=\mathcal{H}_A\otimes\mathcal{H}_B$ of subsystem spaces $\mathcal{H}_A$ and $\mathcal{H}_B$. 

Even if subsystems are identical and have unitarily equivalent spaces $\mathcal{H}_A\propto \mathcal{H}_B \propto \mathcal{H}= \mathcal{C}^N$, the common space of states is $H=\mathcal{H}\otimes \mathcal{H} = \mathcal{C}^{N^2}$. Generally, dimension of the space of the system states $H$ is $N_{sys}=N_A N_B$.

It is significant that system space can be fibred by means of group of symmetry of subsystem $U\left(2\right)$ into direct sum of irreducible representations of that group. Dimension of each irreducible representation takes values up to the sum of subspace dimensions, not the product of those. 

Now we denote as $\hat{\vec{S}}_A$ and  $\hat{\vec{S}}_B$ momentum operators for subsystems $\mathcal{A}$ and $\mathcal{B}$ respectively. So, we can define set of irreducible representations $\mathcal{H}_j$ by rule of addition of angular momentum. Let $N_A - N_B = d\ge 0$, thus subspaces 
$\left\{\mathcal{C}^d,\ \mathcal{C}^{d+1}, \ldots , \mathcal{C}^{N_A + N_B-1} \right\}$ contain the irreducible multiplets.

\subsection{Reconstruction of state}
Now we suppose that we have representative set of measured values reflecting various properties of system and of both of its parts, and we are going to determine if the system is entangled or not. First we examine the set as to its adequacy for full determination of state of each subsystem.

\subsubsection{Observables}
State of the system is determined under conditions that all components $\rho_{m,n}$ of density matrix are given, thus set of measured values  is to be large enough for calculation of all the components.

Process of measurement takes place as count by set of independent detectors. Independence implies that each time only one detector counts. Completeness and purity of detectors is essential as well. Purity implies that projection of a state of the system on each detected state is one-dimensional; independence  -- that these projections are orthogonal,  and completeness -- that these projections give resolution of identity.

In terms of a detector operator $\hat{D}_k$: 
 independence and purity $\hat{D}_k\hat{D}_n=\hat{D}_k\delta_{k,n}$; 
 completeness $\sum_{\forall k}{\hat{D}_k}=\hat{I}$.

Set of measured values is a set of probabilities for each detector
\[p_m=\tr{\hat{D}_m\hat{\rho}}=\brcat{m}{\hat{\rho}}=\rho_{m,m}.
\]
 We can assign to each detector an observable value $O_k$ 
and so we define an observable through its decomposition 
\[\sum_{\forall k}{O_k\hat{D}_k}\stackrel{def}{\rightarrow}\hat{O}.\]

Different sets of observable values define various observables forming a class of commutable observables. Typical example is $\hat{S}_z$ given by \eref{def_z}. 

\subsubsection{Ladder basis}
Any class of commutable observables can be represented as polynomial of typical element of the class. We can represent each such example by power of ladder operators
\begin{equation}\label{def_lad_m}
	\hat{S}_\pm^{\left(m\right)}=\hat{S}_\pm^{m},
\end{equation}
as sum or difference $\hat{S}_+^{\left(m\right)}\pm\hat{S}_-^{\left(m\right)}$ or as $\hat{S}_+^{\left(m\right)}\hat{S}_z\pm\hat{S}_z\hat{S}_-^{\left(m\right)}$. Thus we can describe all sets of observables by two sequences
$\left\{\hat{S}_+^{\left(m\right)}\hat{S}_z\pm\hat{S}_z\hat{S}_-^{\left(m\right)}, m=0\ldots N-1\right\}$ or by real and imaginary parts of the sequence  
$\left\{\hat{S}_+^{\left(m\right)}\hat{S}_z, m=0\ldots N-1\right\}$.

Since each class of commutable observables is represented by a polynomial function there exist $2N^2$ observables 
$\hat{O}_{n,m}=\left(\hat{S}_+^{\left(m\right)}\hat{S}_z\right)^n$
with matrix elements
$O_{n,m}^{p,k}=\bracat{p}{\left(\hat{S}_+^{\left(m\right)}\hat{S}_z\right)^n}{k}$.
 
The values of those observables
$\aver{\hat{O}_{n,m}}=\tr{\hat{O}_{n,m}\hat{\rho}}$
for a given state $\hat{\rho}$ depend on coefficients of density matrix and lead to a linear system of equations for these coefficients
\begin{equation}\label{det_rho}
	\sum_{p,k=1}^{N}{O_{n,m}^{p,k}\rho_{p,k}}=\aver{\hat{O}_{n,m}}.
\end{equation}

\subsection{Measurement of composite system}
Interrelation between both parts of the system brings up correlations of measured values.
We suppose that each count of detector measuring subsystem $\mathcal{A}$ is accompanied by a count of some detector measuring subsystem $\mathcal{B}$. We can interprete each pair of detectors $\hat{D}_n^{\left\{A\right\}}$ and $\hat{D}_m^{\left\{B\right\}}$ as a composite detector $\hat{D}_k^{\left\{sys\right\}}$ with number $k$ being a function  $k\left(m,n\right)$ of numbers of detectors of parts. Hence we can describe the composite system by its set of sequences of counts and obtain its density matrix 

\begin{equation}\label{def_comp}
	\hat{\rho}^{\left\{sys\right\}}=\sum_{\forall k,p}
	\rho^{\left\{sys\right\}}_{k,p}\pcat{k}\odot\pbra{p}.
\end{equation}

\subsubsection{Definition of covariance matrix}
Another way to describe a composite system is to supplement independent describtions of each part with account of correlation between observables of different parts.

Correlation has description by covariance matrix with coefficients obtained as estimation of mutual sampling rate limit 
\begin{equation}\label{cova}
	\left\{\frac{N_{k_A \& m_B}}{N_{full}}-\frac{N_{k_A}}{N_{full}}\frac{N_{m_B}}{N_{full}}\rightarrow c_{k,m}:\ k=1\ldots N_A, m=1\ldots N_B\right\}.
\end{equation}
Here we have coefficients of covariance matrix of observables from different parts only. Such coefficients for observables from one part are not measurable because arise from different series of measurements. Thus counts $N_{k_A \& m_B}$ and $N_{full}$ belong to one common series, each coefficient of covariance matrix \eref{cova} originates from its own series and complete measurement of correlation between two parts of given system needs a complete set of $N_A \cdot N_B$ measurement series.

\subsubsection{Determination of covariance matrix}
For a given state of two-part system covariance matrix is determined by average value of common observable $\hat{S}^{\left\{1\right\}}_{n}\hat{S}^{\left\{2\right\}}_{m}$ being the product of corresponding observables of each part
\begin{equation}\label{cov_det}
	C\left(n,m\right)=
	\aver{
	\left[\hat{S}^{\left\{1\right\}}_{n}-\aver{\hat{S}^{\left\{1\right\}}_{n}}\right]
	\left[\hat{S}^{\left\{2\right\}}_{m}-\aver{\hat{S}^{\left\{2\right\}}_{m}}\right]
	}.
\end{equation}

\subsection{Decomposition of state of composite system}
Description of states of composite system can be performed in composite basis $\left\{\pcat{k}\right\}$, as \eref{def_comp}, and in the basis $\left\{\cat{m}\otimes\ecat{n}\right\}$ of direct product of the subsystem states as well.
Relationship between those bases is similar to the relationship between the basis 
of total angular momentum $\left\{\pcat{j,m_j}\right\}$ and the direct product of the bases of orbital angular momentum and spin $\left\{\cat{l,m}\otimes\ecat{m_s}\right\}$.  
There exists a set of well-known rules of correspondence between states of total angular momentum and states of combinations of orbital momentum and spin
\[
	\pcat{j,m_j}=\sum_{m_l=-l\ldots l}{\sum_{m_s=-\frac{1}{2},\frac{1}{2}}{
	C_{j,m_j;m_l,m_s}\cat{l,m_l}\otimes\ecat{m_s},
	}}
\]
where $C_{j,m_j;m_l,m_sj}$ are Clebsch--Gordan coefficients.

In general case we have similar rules of correspondence for composition of two parts with momenta $l$ and $s\le l$ given by
\begin{equation}\label{clebsh}
	\pcat{j,m_j}=\sum_{m_l=-l\ldots l}{\sum_{m_s=-s\ldots s}{
	C_{j,m_j;m_l, m_s}\cat{l,m_l}\otimes\ecat{s,m_s} .
	}}
\end{equation}

\subsubsection{Invariant states}
Invariance of states of composite system under group $U\left(2\right)$ transformation is realised through diagonalization of density matrix on basis of irreducible representations. Therefore we have representation of density matrix as
\begin{equation}\label{diag}
	\hat{\rho}_{sys}=\sum_{j=j_{min}\ldots j_{max}}{\sum_{m_j=-j\ldots j}{
	\rho_{j,m_j}\proa{j,m_j}
	}}.
\end{equation}
Using decomposition of irreducible states by pure states of subsystems we have representation of the density matrix of whole system as combination of density  matrices of subsystems

\[
\hat{\rho}_{sys}=\]
\[\sum_{j,m_j,m_l,m_s;n_l,n_s}{
	\rho_{j,m_j}
C_{j,m_j;m_l, m_s}C_{j,m_j;n_l, n_s}
\cat{l,m_l}\odot\bra{l,n_l}\otimes\ecat{s,m_s}\odot\ebra{s,n_s}
}
\]
Main result of this decomposition is in representation of density matrices of subsystems obtaned by averaging by states of another subsystem
\begin{equation}
\begin{array}{l}
	\hat{\rho}_{A}=\sum_{m_l}{\left(
	\sum_{j,m_j;m_s}{	\rho_{j,m_j}C_{j,m_j;m_l, m_s}^2}\right)
\roa{l,m_l}}\\
	\hat{\rho}_{B}=\sum_{m_s}{\left(
	\sum_{j,m_j;m_l}{	\rho_{j,m_j}C_{j,m_j;m_l, m_s}^2}\right)
\eroa{s,m_s}}\\
\end{array}
\end{equation}
We see that each irreducible part of density matrix of composite system has its own term in density matrices of subsystems and all these parts are diagonal because of special properties of Clebsch--Gordan coefficients.

\subsubsection{States of subsystems}
Each pure state of whole system being irreducible representation has its own density matrices of each subsystem
\[
\hat{\rho}^{\left\{A\right\}}_{j,m_j}=\sum_{m_l}{\left(
	\sum_{m_s}{C_{j,m_j;m_l, m_s}^2}\right)
\roa{l,m_l}};\] 
\[\hat{\rho}^{\left\{B\right\}}_{j,m_j}
=\sum_{m_s}{\left(
	\sum_{m_l}{	C_{j,m_j;m_l, m_s}^2}\right)
\eroa{s,m_s}}.
\]
We denote

\begin{equation}
	\rho^{\left\{A\right\}}_{j,m_j;m_l}=\sum_{m_s}{C_{j,m_j;m_l, m_s}^2};\ 
	\rho^{\left\{B\right\}}_{j,m_j;m_s}=\sum_{m_l}{C_{j,m_j;m_l, m_s}^2}
\end{equation}
and obtain diagonal representation of density matrices of each subsystem for each pure irreducible state of whole system
\begin{equation}\label{sub_rho}
\hat{\rho}^{\left\{A\right\}}_{j,m_j}=\sum_{m_l}{\rho^{\left\{A\right\}}_{j,m_j;m_l}
\roa{l,m_l}};\ 
\hat{\rho}^{\left\{B\right\}}_{j,m_j}
=\sum_{m_s}{\rho^{\left\{B\right\}}_{j,m_j;m_s}
\eroa{s,m_s}}.
\end{equation}
Almost each one of pure irreducible states of whole system consists of more than one product of states of subsystems like complete angular momentum of electron $m_j$ formed by two orbital substates with orbital momenta $m_j-1/2$ and $m_j+1/2$. Only two extreme states with momenta $m_j=\pm j$ are formed as products of orbital and spin states and only two extreme states with $m_j=\pm\left(l+s\right)$ are states of independent subsystems.

Density matrices \eref{sub_rho} are diagonal in common basis so density matrix of any part for whole mixed system \eref{def_comp}  is diagonal as well:
\[
\hat{\rho}^{\left\{A\right\}}
=\sum_{j,m_j}{\rho^{\left\{sys\right\}}_{j,m_j}\hat{\rho}^{\left\{A\right\}}_{j,m_j}}
=\sum_{m_l}{\sum_{j,m_j}{\rho^{\left\{sys\right\}}_{j,m_j}\rho^{\left\{A\right\}}_{j,m_j;m_l}}
\roa{l,m_l}}
\]
\[
\hat{\rho}^{\left\{B\right\}}
=\sum_{j,m_j}{\rho^{\left\{sys\right\}}_{j,m_j}\hat{\rho}^{\left\{B\right\}}_{j,m_j}}
=\sum_{m_s}{\sum_{j,m_j}{\rho^{\left\{sys\right\}}_{j,m_j}\rho^{\left\{B\right\}}_{j,m_j;m_s}}
\eroa{s,m_s}}
\]
Diagonal elements of these density matrices are

\begin{equation}\label{sub_diagA}
\rho^{\left\{A\right\}}_{m_l}=\sum_{j,m_j}{\rho^{\left\{sys\right\}}_{j,m_j}\rho^{\left\{A\right\}}_{j,m_j;m_l}};
\end{equation}
\begin{equation}\label{sub_diagB}
\rho^{\left\{B\right\}}_{m_s}=\sum_{j,m_j}{\rho^{\left\{sys\right\}}_{j,m_j}\rho^{\left\{B\right\}}_{j,m_j;m_s}}
\end{equation}
They are equal only in the case of same dimensions of state spaces of both parts. 

Observables $\hat{L}_z$,  $\hat{S}_z$ and $\hat{J}_z$ are measurable jointly, so joint probabilities of their measurement exist.  
\section{Entropy}
Information that can be obtained in measurement of a system is given by Shannon entropy and is limited from above by von Neumann entropy:
\[S_S=-\sum_{\forall k}{p_k\log_2 p_k};\ S_N=-\tr{\hat{\rho}\log_2\hat{\rho}}.
\]
While the space of system states has finite dimensionality, basis diagonalizing density matrix always exists. Thus it is not needed to distinguish between Shannon and von Neumann entropies. The basis $\left\{\cat{k},\; \forall k\right\}$ is composed of eigenvectors of density matrix $\hat{\rho}\cat{k}=p_k\cat{k}$ and gives probabilities $p_k$ as averages $\aver{\hat{D}_k}$ of detectors $\hat{D}_k=\roa{k}$.

\subsection{Entropy of whole system}
Existence of group of symmetry of each subsystem of given system leads to fibering of space of states of the whole system to direct sum of subspaces containing irreducible representations of the group. In addition density matrix of the given system must be diagonal in respective basis
\[\hat{\rho}=\sum_{j,m_j}{\rho_{j,m_j}\proa{j,m_j}},
\]
thus von Neumann entropy of whole system is equal to Shannon entropy.
\[S_{sys}=S_N=-\sum_{j,m_j}{\rho_{j,m_j}\log_2 \rho_{j,m_j}}.
\]
\subsection{Entropies of subsystems of pure system}
For pure system with density matrix $\hat{\rho}=\proa{j,m_j}$ states of each subsystem \eref{sub_rho} are mixed and have equal entropies given by
\[S^{\left\{P\right\}}_{j,m_j}
=-\sum_{m_l}{\rho^{\left\{A\right\}}_{j,m_j;m_l}\log_2 \rho^{\left\{A\right\}}_{j,m_j;m_l}}
=-\sum_{m_s}{\rho^{\left\{B\right\}}_{j,m_j;m_s}\log_2 \rho^{\left\{B\right\}}_{j,m_j;m_s}}
\]
To this expression the name of entropy of entanglement is given in \cite{VP,RHMH} since it has the meaning of entropy produced by disentangling of entangled system. Values of entropy of both subsystems are equal even if spaces of subsystems have different dimensions.

\subsection{Entropies of subsystems of mixed system}
Now we can obtain the entropies for each of subsystems of given mixed system by means of diagonal coefficients \eref{sub_diagA} and \eref{sub_diagB} of respective density matrices
\begin{equation}\label{def_sp}
	S^{\left\{A\right\}}=-\sum_{m_l}{
	\rho^{\left\{A\right\}}_{m_l}\log_2 \rho^{\left\{A\right\}}_{m_l}};\ 
	S^{\left\{B\right\}}=-\sum_{m_s}{
	\rho^{\left\{B\right\}}_{m_s}\log_2 \rho^{\left\{B\right\}}_{m_s}}
	\end{equation}
Substitution of elements of density matrices leads to
\begin{equation}
	\begin{array}{l}
S^{\left\{A\right\}}
=-\sum_{j,m_j}{\rho^{\left\{sys\right\}}_{j,m_j}
\sum_{m_l}{
\rho^{\left\{A\right\}}_{j,m_j;m_l}
\log_2 \sum_{J,m_J}{\rho^{\left\{sys\right\}}_{J,m_J}
\rho^{\left\{A\right\}}_{J,m_J;m_l}}}}	
\\
S^{\left\{B\right\}}
=-\sum_{j,m_j}{\rho^{\left\{sys\right\}}_{j,m_j}
\sum_{m_s}{
\rho^{\left\{B\right\}}_{j,m_j;m_s}
\log_2 \sum_{J,m_J}{\rho^{\left\{sys\right\}}_{J,m_J}
\rho^{\left\{B\right\}}_{J,m_J;m_s}}}}
	\end{array}
\end{equation}
With account of inequality 
\[-\sum_{m_s}{
\rho^{\left\{P\right\}}_{j,m_j;m_s}
\log_2 \sum_{J,m_J}{\rho^{\left\{sys\right\}}_{J,m_J}
\rho^{\left\{P\right\}}_{J,m_J;m_s}}} \ge 
-\sum_{m_s}{
\rho^{\left\{P\right\}}_{j,m_j;m_s}
\log_2 \rho^{\left\{P\right\}}_{j,m_j;m_s}}
=S^{\left\{P\right\}}_{j,m_j}
\]
we obtain inequalities giving the lower bounds of entropies of subsystems
\begin{equation}
	\begin{array}{l}
S^{\left\{A\right\}}\ge
\sum_{j,m_j}{\rho^{\left\{sys\right\}}_{j,m_j}S^{\left\{A\right\}}_{j,m_j}};
\\
S^{\left\{B\right\}}
\ge\sum_{j,m_j}{\rho^{\left\{sys\right\}}_{j,m_j}S^{\left\{B\right\}}_{j,m_j}}.
	\end{array}
\end{equation}
Similarly, entropy of each subsystem has as its lower bound the entropy of whole system
\begin{equation}
	S^{\left\{A,B\right\}}\ge 
	-\sum_{j,m_j}{
	\rho^{\left\{sys\right\}}_{j,m_j}\log_2 \rho^{\left\{sys\right\}}_{j,m_j}}
	=S^{\left\{sys\right\}}.
\end{equation}
Upper bounds of entropies of subsystems result from the finite dimensionalities of spaces of states of subsystems and are equal to $\log_2 N_{A,B}$.

\subsection{Production of entropy}
Difference between density matrix of the whole system and direct product of density matrices of its subsystems leads to difference between entropy of whole system and sum of entropies of the subsystems. 
Process of measurement of the subsystems, of one or both, divides the system into two parts, thus entropy of system turns to sum of entropies of the subsystems. The sum is always larger than initial entropy so process of measurement of any part of a composite system produces entropy of the system.

Resulting entropy of the subsystem remains smaller than the entropy of whole system or equal to it if the system is mixed. Only in the case of whole system or its part being entangled resulting entropy exceeds entropy of the whole system.

Thus, excess of entropy of a subsystem over entropy of the whole system indicates the presence of entaglement in the system.

\section{Qubit and qutrit entanglement}
Let the system has two nonequivalent parts - qubit and qutrit. 

 Subsystem $A$ is similar to angular subsystem with angular momentum $l=1$, and subsystem $B$ is similar to subsystem of electron spin. Space of states of whole system has dimension $3\otimes 2 =6 =2+4$. 

Model of spin-orbit coupling provides physical interpretation of bases being eigenvectors of $\hat{L}_z$ and $\hat{s}_z$ operators.
\paragraph{States of subsystems}
Nondiagonal elements of part $\mathcal{A}$ are present in pairs with the nondiagonal elements of part $\mathcal{B}$ only, so averaging by states of one part leads to state of another part with diagonal elements only.
\begin{figure}[ht!]
\begin{center}
\includegraphics[scale=0.6]{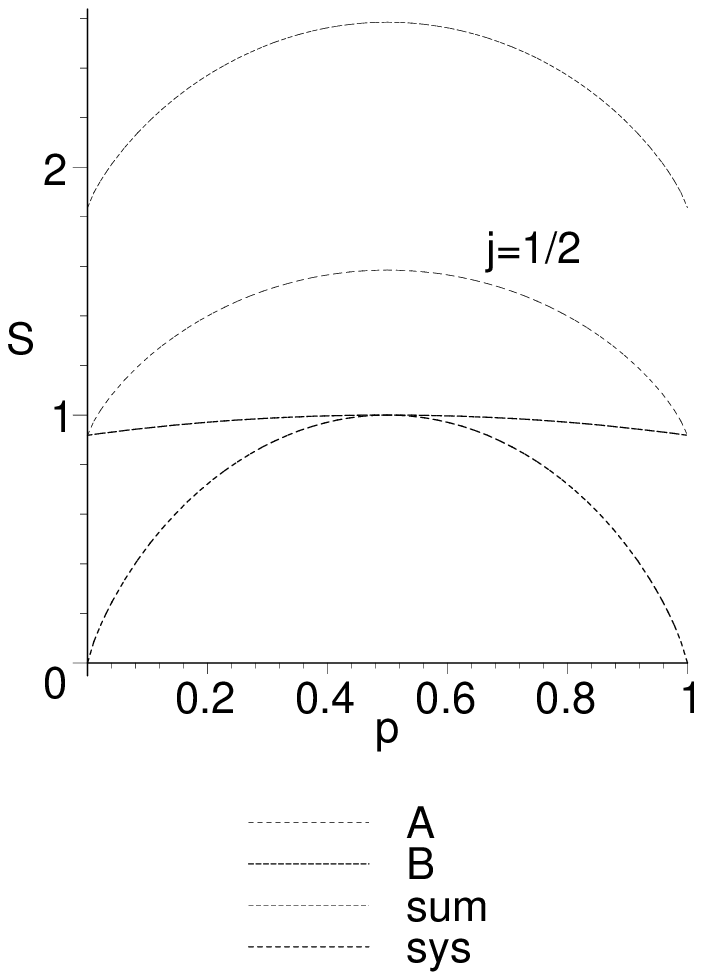}\includegraphics[scale=0.6]{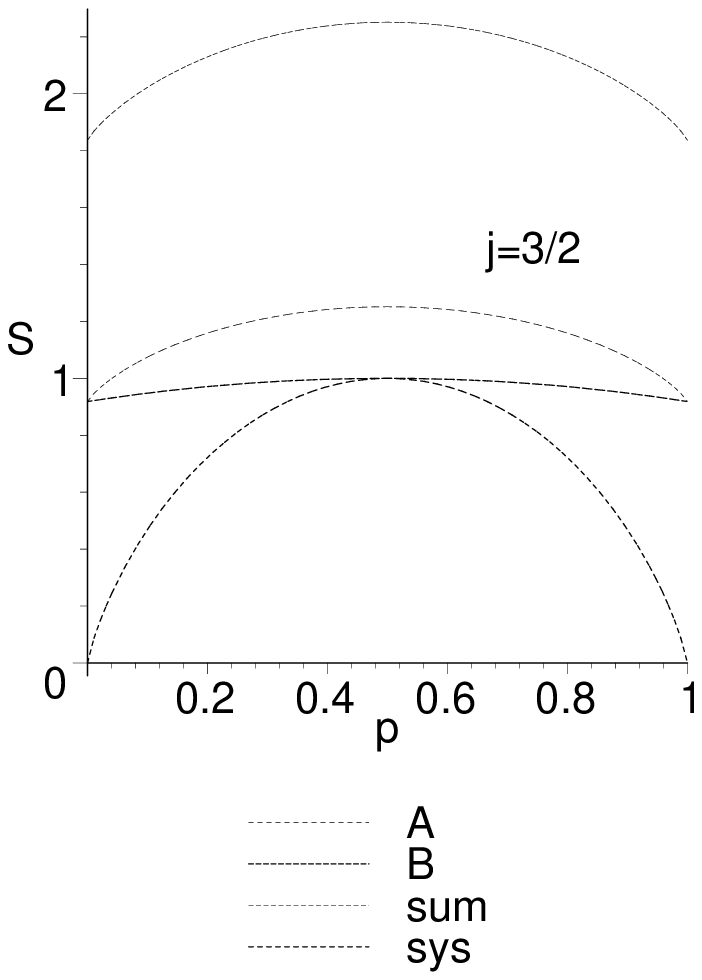}
\caption{Entropy of entangled system "{\it sys}" and its subsystems {\it A,B}. Left graph shows mixing of states $\pcat{\frac{1}{2},\pm\frac{1}{2}}$, right -- $\pcat{\frac{3}{2},\pm\frac{1}{2}}$.}\label{fig1}
\end{center}	
\end{figure}

Entropies of subsystems for pure states 
$S_{\frac{3}{2},\pm\frac{3}{2}}^{\left\{1\right\}}=S_{\frac{3}{2},\pm\frac{3}{2}}^{\left\{\frac{1}{2}\right\}}
=0.$
$S_{\frac{1}{2},\pm\frac{1}{2}}^{\left\{1\right\}}=S_{\frac{1}{2},\pm\frac{1}{2}}^{\left\{\frac{1}{2}\right\}}
=S_{\frac{3}{2},\pm\frac{1}{2}}^{\left\{1\right\}}=S_{\frac{3}{2},\pm\frac{1}{2}}^{\left\{\frac{1}{2}\right\}}
=\sth{\frac{1}{3}}\approx0.918$.

Six parameters of mixed states $\rho_{\frac{3}{2},\pm\frac{3}{2}}$, $\rho_{\frac{3}{2},\pm\frac{1}{2}}$, $\rho_{\frac{1}{2},\pm\frac{1}{2}}$. denote probabilities of each given pure state.

Entropies of subsystems $S^{\left\{\frac{1}{2}\right\}}=\sth{p}$ and $S^{\left\{1\right\}}=\sthh{p_-}\sthh{p_0}\sthh{p_+}$ 
 depend on given probabilities by means of cumulation
$ p=\frac{2}{3}\rho_{\frac{1}{2},+\frac{1}{2}}
+\frac{1}{3}\rho_{\frac{1}{2},-\frac{1}{2}}
+\frac{1}{3}\rho_{\frac{3}{2},+\frac{1}{2}}
+\frac{2}{3}\rho_{\frac{3}{2},-\frac{1}{2}}
+\rho_{\frac{3}{2},-\frac{3}{2}}$ and 
$ p_-=\frac{2}{3}\rho_{\frac{1}{2},-\frac{1}{2}}
+\frac{1}{3}\rho_{\frac{3}{2},-\frac{1}{2}}
+\rho_{\frac{3}{2},-\frac{3}{2}}$, 
$p_0=\frac{1}{3}\rho_{\frac{1}{2},+\frac{1}{2}}
+\frac{1}{3}\rho_{\frac{1}{2},-\frac{1}{2}}
+\frac{2}{3}\rho_{\frac{3}{2},+\frac{1}{2}}
+\frac{2}{3}\rho_{\frac{3}{2},-\frac{1}{2}}$, 
$p_+=\frac{2}{3}\rho_{\frac{1}{2},+\frac{1}{2}}
+\frac{1}{3}\rho_{\frac{3}{2},+\frac{1}{2}}
+\rho_{\frac{3}{2},+\frac{3}{2}}$

\section{Summary}
\begin{itemize}
	\item 
Invariance to group $U(2)$ is essential property of an entangled state.
\item
Process of measurement of any part of a composite system produces entropy of the system and creates an information about the system.
\item
Excess of entropy of a subsystem over entropy of the whole system indicates the presence of entaglement in the system.
\end{itemize}

\end{document}